\documentclass[aps,preprintnumbers,nofootinbib,prd,superscriptaddress,tightenlines,notitlepage,twocolumn,showpacs,floatfix]{revtex4-1}
\usepackage[colorlinks]{hyperref}
\usepackage{tabularx}
\usepackage{graphicx}
\usepackage{amssymb,bm,tensor}
\usepackage{textcomp}
\usepackage{xcolor}
\usepackage{amsmath}
\usepackage[varg]{txfonts}
\usepackage{enumerate}
\usepackage{mathtools}
\usepackage[normalem]{ulem}
\usepackage{bm} 
\usepackage[OT1]{fontenc}
\usepackage{aas_macros}

\newcommand{\rmd}{{\rm d}}
\newcommand{\hata}{\hat{a}}
\newcommand{\hatb}{\hat{b}}
\newcommand{\hatc}{\hat{c}}
\newcommand{\hatn}{\hat{n}}

\newcommand{\KIAA}{\affiliation{Kavli Institute for Astronomy and Astrophysics,
Peking University, Beijing 100871, China}}
\newcommand{\MPIfR}{\affiliation{Max-Planck-Institut f\"ur Radioastronomie, Auf
dem H\"ugel 69, D-53121 Bonn, Germany}}
\newcommand{\ERAU}{\affiliation{Department of Physics and Astronomy, Embry-Riddle Aeronautical
University, Prescott, Arizona 86301, USA}}

\begin{document}

\title{Testing the Gravitational Weak Equivalence Principle in the
Standard-Model Extension \\ with Binary Pulsars}
\date{\today}
\author{Lijing Shao}\email{lshao@pku.edu.cn}\KIAA\MPIfR
\author{Quentin G. Bailey}\email{baileyq@erau.edu}\ERAU

\begin{abstract}
  The Standard-Model Extension provides a framework to systematically investigate
  possible violation of the Lorentz symmetry. Concerning gravity, the
  linearized version was extensively examined. We here cast the first set of
  experimental bounds on the nonlinear terms in the field equation from the
  anisotropic cubic curvature couplings. These terms introduce body-dependent
  accelerations for self-gravitating objects, thus violating the
  gravitational weak equivalence principle (GWEP). Novel phenomena, that are absent in
  the linearized gravity, remain experimentally unexplored. 
  We constrain them with precise binary-orbit measurements from pulsar
  timing, wherein the high density and large compactness of neutron stars are
  crucial for the test. It is the first study that seeks
    GWEP-violating signals in a fully anisotropic framework with Lorentz
  violation.
\end{abstract}

\maketitle


\section{Introduction}
\label{sec:intro}

Einstein's general relativity (GR) is considered as {\it tour de force} in
describing gravity~\cite{Einstein:1915ca, Misner:1974qy}. For the past more
than 100 years, GR has passed numerous experimental tests with flying
colors~\cite{Will:2014kxa, Kostelecky:2008ts, Wex:2014nva, Berti:2015itd,
Shao:2016ezh, Tasson:2014dfa}. The wisdom in GR is concisely condensed into
the Einstein-Hilbert Lagrangian, ${\cal L}_{\rm EH} = \sqrt{-g}\left(
R-2\Lambda \right)/ 16\pi G$, where $g$ is the determinant of the metric
$g_{\mu\nu}$, $R$ is the Ricci scalar, and $\Lambda$ is the cosmological
constant. Einstein's field equations are derived through a variation of ${\cal
L}_{\rm EH}$ with respect to $g_{\mu\nu}$. The inherent local Lorentz
invariance (LLI) and diffeomorphism symmetry are essential properties for GR
from a theoretical viewpoint~\cite{Kostelecky:2017zob, Misner:1974qy}.

Open questions in the contemporary modern physics, like the very nature of
dark matter and dark energy, encourage us to test the underlying fundamental
principles in GR. LLI is one of the most important. In a flat
spacetime, it links the inertial frames that are relatively moving with
respect to each other, while in the curved space, it addresses the property
of the tangent space at every single point~\cite{Misner:1974qy}. However,
from a deeper understanding LLI may not be ``god-given''~\cite{Witten:2017hdv},
as in string theory~\cite{Kostelecky:1988zi, Kostelecky:1991ak} and loop
quantum gravity~\cite{Gambini:1998it}. Experimental examination, that might
either strengthen further our confidence in GR or lead to discoveries beyond
the current paradigm, is vital.

The Standard-Model Extension (SME) provides an effective field theory (EFT) that
extends our currently well adopted field theories of GR and particle physics.
It incorporates all Lorentz-violating (LV) operators which are made out of matter
fields and the metric field
$g_{\mu\nu}$~\cite{Colladay:1996iz, Colladay:1998fq, Kostelecky:2003fs}. We
call the covariant coupling coefficients the {\it coefficient fields}. In
this {\it Letter}, we focus on the gravity sector of the
SME~\cite{Kostelecky:2003fs, Bailey:2006fd, Bailey:2014bta, Bailey:2016ezm,
Bailey:2017lbo}, and leave LV matter-gravity
couplings~\cite{Kostelecky:2010ze, Jennings:2015vma} for a
future study.

In the SME, conventionally LV operators are sorted according to their
mass dimensions as {\it per} EFT~\cite{Kostelecky:1994rn, Weinberg:2009bg}. In
general, operators with higher mass dimensions are believed to be suppressed.
At mass dimension 4, extra coefficient fields in the gravity sector have in
total 20 degrees of freedom.  Only 9 of them enter the leading-order
post-Newtonian (PN) dynamics~\cite{Bailey:2006fd}. They are well constrained by
various experiments, including lunar laser ranging~\cite{Battat:2007uh,
Bourgoin:2017fpo}, atom interferometers~\cite{Muller:2007es}, cosmic
rays~\cite{Kostelecky:2015dpa}, pulsar timing~\cite{Shao:2014oha,
Shao:2014bfa}, planetary orbital dynamics~\cite{Hees:2015mga},
super-conducting gravimeters~\cite{Flowers:2016ctv, Shao:2017bgz}, and
gravitational waves~\cite{Monitor:2017mdv} (see \citet{Hees:2016lyw} for
review). Then at mass dimension 5, new operators introduce a CPT-violating
gravitational force. Gravitational waves~\cite{Kostelecky:2016kfm} and binary
pulsars~\cite{Shao:2018vul} were used to put bounds. At mass dimension 6 and
higher, short-range experiments in laboratories are extremely powerful to cast
useful limits~\cite{Bailey:2014bta, Shao:2016cjk, Kostelecky:2016uex,
Shao:2018lsx}.  We refer the reader to the updated data
tables~\cite{Kostelecky:2008ts} for details.

Nevertheless, all limits mentioned above are based on the linearized version
of gravity~\cite{Kostelecky:2017zob}, where terms are only kept up to the
quadratic order of the metric perturbation $h_{\mu\nu} \equiv g_{\mu\nu} -
\eta_{\mu\nu}$ in the Lagrangian with $\eta_{\mu\nu}$ being the flat-spacetime metric. 
After variation with respect to
$h_{\mu\nu}$, only linear terms appear in the field equations. The linearized
gravity limit unavoidably misses interesting features that only come from nonlinear
terms. \citet{Bailey:2016ezm} has analyzed an interesting example where the
Lagrangian contains a term proportional to the cubic power of the Riemannian tensor
$R_{\mu\nu\rho\lambda}$. When restricting to the cubic order of $h_{\mu\nu}$
in the Lagrangian (hence, quadratic order in the field
equations), body-dependent effects appear for self-gravitating objects. This
is a novel effect for the gravity sector, and it reminds us the
well-known {\it Nordtvedt effect}~\cite{Nordtvedt:1968qs, Will:1993ns} that
is closely related to the gravitational weak equivalence principle
(GWEP)~\cite{Will:1993ns, Will:2014kxa, Shao:2016ezh}. In this {\it Letter} we
analyze this phenomenon in detail with precision binary orbits from pulsar
timing~\cite{Lorimer:2005misc, Wex:2014nva, Shao:2014wja}, and put the first
set of observational bounds.

Unless otherwise stated, we use units where $c=1$.

\section{Anisotropic cubic curvature couplings}
\label{sec:th}

The gravity sector of the SME was built to extend GR by including coefficient
fields that couple to the metric and its
derivatives~\cite{Kostelecky:2003fs}. LLI is broken by the cosmological
condensation of these coefficient fields. In the 4-dimensional (4d)
Riemann-Cartan spacetime, to be compatible with geometrical identities, the
breaking should be {\it spontaneous}, instead of {\it
explicit}~\cite{Kostelecky:2003fs, Bluhm:2004ep}. It is set by minimizing the energy of
coefficient fields through a Higgs-like mechanism. Nevertheless, unlike the
Higgs, coefficient fields can take spacetime indices, thus their nonzero
vacuum expectation values (VEVs) break the Lorentz symmetry. In other words,
while physical states are LV, the underlying fundamental
theory is Lorentz-invariant~\cite{Tasson:2016xib}.

Being general, the Lagrangian in the SME~\cite{Kostelecky:2003fs} reads
${\cal L} = {\cal L}_{\rm EH} + {\cal L}_{\rm LV} + {\cal L}_{\rm k} + {\cal
L}_{\rm m}$, where (i) ${\cal L}_{\rm m}$ is the matter sector; (ii) ${\cal
L}_{\rm k}$ describes the dynamics (including the symmetry breaking) of the
coefficient fields, whose details are not crucial here; (iii) ${\cal L}_{\rm
LV}$ contains the couplings between the LV coefficient fields
with the gravitational field. 
The terms in ${\cal
L}_{\rm k}$ can be organized using the mass dimension of the curvature operator they contain~\cite{Kostelecky:1994rn,Kostelecky:2003fs,Bailey:2014bta}.
As an interesting case study, we focus on
operators of mass dimension 8~\cite{Bailey:2016ezm},
\begin{align}
  {\cal L}^{(8)}_{\rm LV} &= \frac{\sqrt{-g}}{16\pi G}
  k^{(8)}_{\alpha\beta\gamma\delta\kappa\lambda\mu\nu\epsilon\zeta\eta\theta}
  R^{\alpha\beta\gamma\delta} R^{\kappa\lambda\mu\nu}
  R^{\epsilon\zeta\eta\theta}\,.
  \label{eq:L8}
\end{align}
Here
$k^{(8)}_{\alpha\beta\gamma\delta\kappa\lambda\mu\nu\epsilon\zeta\eta\theta}$
are the coefficient fields, and they have physical dimensions of quartic
length. 
Like other coefficients, 
they can be composite of lower-order tensors, as was shown, for example, 
in Sec.~IV of Ref.~\cite{Bailey:2006fd} with bumblebee models.
We use the compact grouping notation~\cite{Bailey:2016ezm}, as $k^{(8)}_{\cal ABC} \equiv
k^{(8)}_{\alpha\beta\gamma\delta\kappa\lambda\mu\nu\epsilon\zeta\eta\theta}$,
$R^{\cal A} R_{\cal A} \equiv R^{\alpha\beta\gamma\delta}
R_{\alpha\beta\gamma\delta}$, and so on. By design, when $\left| h_{\mu\nu} \right| \ll
1$, $R^{\alpha\beta\gamma\delta} \sim {\cal O} \left( h \right)$, and ${\cal
L}^{(8)}_{\rm LV} \sim {\cal O} \left( h^3 \right)$. 
Therefore, the field
equations have no ${\cal O} \left( h \right)$ contributions from Eq.~(\ref{eq:L8}). 
Other possible 8d terms proportional to ${\cal D}^\alpha {\cal D}^\beta R^{\cal A}
R^{\cal B}$ and ${\cal D}^\alpha {\cal D}^\beta {\cal D}^\gamma {\cal
D}^\delta R^{\cal A}$ introduce  lower-order contributions of $h$.
Therefore, in the sense of solely studying nonlinear terms and body-dependent effects in the gravity sector of the SME, 
${\cal L}^{(8)}_{\rm LV}$ is complete at leading
order,
saving for possible contributions from the dynamical terms in ${\cal L}_{\rm k}$.

 Through symmetry breaking, $k^{(8)}_{\cal ABC}$ obtains its VEV,
 $\bar{k}^{(8)}_{\cal ABC}$. In principle we still need to account for the
 dynamics of the fluctuation $\tilde{k}^{(8)}_{\cal ABC} \equiv k^{(8)}_{\cal
 ABC} - \bar k^{(8)}_{\cal ABC}$ to be fully compatible with the geometry.
 However if we restrict to ${\cal O}\left( h^2 \right)$ terms in the field
 equation, $\tilde{k}^{(8)}_{\cal ABC}$ does not enter~\cite{Bailey:2006fd,
 Kostelecky:2010ze, Bailey:2014bta}. Then, after imposing $\partial_\mu \bar
 k^{(8)}_{\cal ABC} = 0$ in an asymptotically flat Cartesian coordinate, the
 field equation simply reads~\cite{Bailey:2016ezm},
\begin{align} \label{eq:field:2}
  G_{\mu\nu} &= 8\pi G \left( T^{\rm m}_{\mu\nu} +T^{\rm
  k}_{\mu\nu}\right) + 6 \bar{k}^{(8)}_{\alpha\mu\nu\beta {\cal AB}}
  \partial^\alpha \partial^\beta \left( R^{\cal A} R^{\cal B} \right)  + {\cal O} \left( h^3 \right) \,,
\end{align}
where $T^{\rm m}_{\mu\nu}$ and $T^{\rm k}_{\mu\nu}$ are the
energy-momentum tensors from ${\cal L}_{\rm m}$ and ${\cal L}_{\rm k}$
respectively.  Under mild assumptions on the nature of the
  dynamical terms for the coefficients $\tilde{k}^{(8)}_{\cal ABC}$,
  we hereafter neglect the contributions from the stress-energy tensor
  $T^{\rm k}_{\mu\nu}$.  For details on this assumption see section II
  in Ref.~\cite{Bailey:2016ezm} and
Refs.~\cite{Seifert:2009vm,Altschul:2010as}.

\section{Binary pulsars}
\label{sec:psr}

Using the technique of PN calculations~\cite{Will:1993ns, Bailey:2006fd},
it was shown that~\cite{Bailey:2016ezm}, from Eq.~(\ref{eq:field:2}), the only
correction to the PN metric in GR is $\delta h_{00}$ in $h_{00}$ at ${\cal
O}\left( v^4/c^4 \right)$. It satisfies a Poisson-like equation~\cite{Bailey:2016ezm},
\begin{equation}
  \label{eq:deltah00}
  \nabla^2 \delta h_{00} = -96 \left( \bar k^{(8)}_{\rm eff} \right)_{jklmnp} \partial_j \partial_k \left( \partial_l \partial_m U \, \partial_n \partial_p U \right) \,,
\end{equation}
where $U$ is the Newtonian potential, and $\bar k^{(8)}_{\rm eff}$, with 56 independent components, are linear combinations of $\bar k^{(8)}_{\cal ABC}$; see Eq.~(17) in Ref.\,\cite{Bailey:2016ezm}.

Consider a binary is composed of bodies
$a$ and $b$ with masses $m_a$ and $m_b$. The
acceleration for body $a$ is, $\bm{a}_a \equiv \rmd^2 \bm{r}_a / \rmd t^2 = - G m_b
\hat{\bm{n}} / r^2 + \bm{a}_a^{\rm PN} + \delta \bm{a}_a$, where $\bm{r}
\equiv \bm{r}_a - \bm{r}_b$ is the separation vector, and $r
\equiv \left| \bm{r} \right|$, $\hat{\bm{n}} \equiv \bm{r}/r$. Interchanging 
indices $a \leftrightarrow b$ gives the acceleration for $b$. The first two terms give the Newtonian and PN accelerations in
GR, while the last term is the abnormal acceleration from Eq.~(\ref{eq:L8}).

The novel aspect of the abnormal acceleration comes from the dependence on
quantities $P_a$, $\tilde{P}_a$, and
$\tilde{P}_a^\prime$~\cite{Bailey:2016ezm},
\begin{align}
  P_a &\equiv \frac{1}{m_a} \int_a \rmd^3 \bm{r} \rho_a^2 \,, \\
  \tilde{P}_a &\equiv \frac{1}{35 m_a} \left( 8\pi \int_a \rmd^3
  \bm{r} \rho_a^2 r^2 + 46 \frac{\Omega_a}{G} \right) \,, \\
  \tilde{P}_a^\prime &\equiv \frac{1}{35 m_a} \left( 16\pi \int_a
  \rmd^3 \bm{r} \rho_a^2 r^2 - 48 \frac{\Omega_a}{G} \right) \,,
\end{align}
where $\rho_a$ and $\Omega_a$ are the density and the (Newtonian)
gravitational self-energy of body $a$, respectively.
The dependence on the internal
structure of objects is a distinct feature due to the nonlinear
terms coupled to $\bar k^{(8)}_{\cal ABC}$'s  VEVs~\cite{Bailey:2016ezm}. It is independent of the multipole structure,
persisting even in the limit of vanishing multipole moments and
tidal forces for perfectly spherical objects. Therefore such a dependence
violates the GWEP and generalizes the well-known Nordtvedt
effect~\cite{Nordtvedt:1968qs, Will:1993ns, Poisson:2014misc, Shao:2016ezh}.
It is absent in the linearized gravity~\cite{Bailey:2006fd,
Kostelecky:2017zob}. This interesting feature is our major motivation to
study the Lagrangian (\ref{eq:L8}).

Roughly speaking, with a uniform density one has $P_a \sim \rho_a$ and
$\tilde{P}_a \sim \tilde{P}_a^\prime \sim m_a / R_a$ where $R_a$ is the
radius of body $a$. The denser of the body (or, the more compact of the
body), the larger of these quantities. Pulsars, with their extremely dense
nuclear matters and significant compactnesses, fit into this scenario
ideally. It is easy to verify that, in a binary system, the denser object dominates the abnormal acceleration. For example, for
neutron star--white dwarf (NS-WD) binaries, we only need to consider the
GWEP-violating contribution from NSs since WDs are weak-field objects with
$\rho_{\rm WD} \ll \rho_{\rm NS}$.

For NSs, the $P_a$ contribution is dominant over $\tilde{P}_a$ (and $\tilde{P}_a^\prime$), by $ \sim\left(
G^2 m_b P_a / r^4 \right) / \left( {G^2 m_b \tilde P_a}/{r^6} \right) \sim {r^2}/{R_a^2} \gg 1$. Thus we only need to
consider the anomalous acceleration $\propto P_a/r^{-4}$. This simplification is
not valid for laboratory short-range gravity tests where the separation of
bodies is comparable to the size of the objects, and the full Eq.~(\ref{eq:deltah00}) is needed~\cite{Bailey:2016ezm,
Shao:2018lsx}.

At the first-order approximation, we assume that the bodies have a
uniform density, which introduces a difference $\lesssim 20\%$
in $P_a$ with respect to a more realistic density profile.  We define
an effective radius $\bar R$ via, $M / \bar R^3 \equiv m_a / R_a^3 +
m_b / R_b^3$ where $M \equiv m_a + m_b$. Thus, we have $\bar R \approx
R_{\rm NS}$ for NS-NS binaries, and $\bar R \approx \left( {M}/{m_{\rm
NS}} \right)^{1/3} R_{\rm NS}$ for NS-WD binaries.  Then the abnormal
{\it relative} acceleration reads,
\begin{align} 
  \delta a^j \equiv \delta a^j_a - \delta a^j_b \approx -432
  \frac{\left( GM \right)^2}{\bar R^3} \frac{K_{kl} \hatn^k \hatn^l \hatn^j -
  \frac{2}{5} K_{jk} \hatn^k}{r^4} \,, \label{eq:del:relative:acc}
\end{align}
where $K_{jk}$ are linear combinations of $\bar k^{(8)}_{\rm
eff}$ given in Eq.~(37) in Ref.\,\cite{Bailey:2016ezm}. It is a symmetric traceless tensor with 5 independent degrees of freedom.  $\delta a^j$
resembles an effective anisotropic quadrupole moment. However, the Newtonian
quadrupole moment decreases when the body gets more compact,
while the GWEP-violating effect has the opposite behavior~\cite{Bailey:2016ezm}.

With osculating elements from celestial mechanics~\cite{Poisson:2014misc},
\citet{Bailey:2016ezm} obtained from Eq.~(\ref{eq:del:relative:acc}) the
secular changes of orbital elements averaging over a Keplerian orbital period $P_b$,
$\left\langle {\rmd a}/{\rmd t} \right\rangle =
\left\langle {\rmd e}/{\rmd t} \right\rangle = 0$, and,
\begin{align}
  \left\langle {\rmd i}/{\rmd t} \right\rangle &= 2 {\cal F} \left(
  K_{\hata\hatc} \cos\omega - K_{\hatb\hatc} \sin\omega \right) \,,
  \label{eq:didt} \\
  \left\langle {\rmd \omega}/{\rmd t} \right\rangle &=
  {\cal F} \left[ {\cal K} - 2 \cot i \left( K_{\hata\hatc} \sin
\omega + K_{\hatb\hatc} \cos\omega \right)\right] \,, \label{eq:domegadt} \\
  \left\langle {\rmd \Omega}/{\rmd t} \right\rangle &= 
    2 {\cal F} \csc i \left( K_{\hata\hatc} \sin\omega + K_{\hatb\hatc}\cos\omega
\right) \label{eq:dOmegadt} \,,
\end{align}
where we have defined,
\begin{align}
  {\cal K} &\equiv K_{\hata\hata} + K_{\hatb\hatb} -2K_{\hatc\hatc}
  \,, \\
  {\cal F} &\equiv \frac{216}{5}  \frac{n_b^3 a}{\bar R^3 \left( 1-e^2
  \right)^2}  \,.
\end{align}
In the above equations, 
$a$ is the relative semimajor axis, $e$ is the eccentricity, $i$ is
the orbital inclination, $\omega$ is the longitude of periastron,
$\Omega$ is the longitude of ascending node, and $n_b \equiv 2\pi /
P_b$~\cite{Will:1993ns, Poisson:2014misc}.  $K_{jk}$ is projected onto
the coordinate frame $\left( \hat{\bm{a}}, \hat{\bm{b}}, \hat{\bm{c}}
\right)$ attached to the pulsar orbit (see Figure~1 in
Ref.\,\cite{Shao:2014bfa}). The formulae for projections can be found
in Eqs.~(18--24) in Ref.\,\cite{Shao:2014bfa}.

As in previous work, the change in the orbital inclination is
converted to the time derivative of a timing parameter $x_p$,
\begin{align}
  \left\langle {\dot x_p}/{x_p} \right\rangle &= 2 {\cal F} \cot i \left(
  K_{\hata\hatc} \cos\omega - K_{\hatb\hatc} \sin\omega \right) \label{eq:xdot}
  \,,
\end{align}
where $x_p \equiv a_p \sin i/c$ is the projected semimajor axis for the pulsar
orbit with $a_p \simeq m_b a/M$. We in general do not measure the longitude of
ascending node $\Omega$ in pulsar timing unless the pulsar is very nearby~\cite{Lorimer:2005misc}, therefore, we will use the
``$\dot\omega$-test'' in Eq.~(\ref{eq:domegadt}) and the ``$\dot x_p$-test'' in
Eq.~(\ref{eq:xdot}) for gravity tests.

From the definition of ${\cal F}$, relativistic binaries with tight orbits
(larger $n_b^3 a$) are preferred to the tests; eccentricity increases ${\cal
F}$ mildly. We have used a handful of well-timed relativistic binary pulsars
to test the CPT-violating gravity~\cite{Shao:2018vul}. Details for these
pulsars are provided collectively in Tables I--III in
Ref.\,\cite{Shao:2018vul}. This collection serves the study here very well.
We divide them into 2 groups: (1) the NS-NS group including 4 systems with
$P_b \lesssim 1$\,day: PSRs~B1913+16~\cite{Weisberg:2016jye},
B1534+12~\cite{Fonseca:2014qla}, B2127+11C~\cite{Jacoby:2006dy}, and
J0737$-$3039A~\cite{Kramer:2006nb}; (2) the NS-WD group including 7 systems
with $P_b \lesssim 2$\,day: PSRs~J0348+0432~\cite{Antoniadis:2013pzd},
J1738+0333~\cite{Freire:2012mg}, J1012+5307~\cite{Lazaridis:2009kq},
J0751$+$1807~\cite{Desvignes:2016yex}, J1802$-$2124~\cite{Ferdman:2010rk},
J1909$-$3744~\cite{Desvignes:2016yex}, and
J2043+1711~\cite{Arzoumanian:2017puf}. The two groups are handled
accordingly. The spread in sky location is important to break the parameter
degeneracy in the tests, as in the earlier work~\cite{Shao:2014oha,
Shao:2014bfa, Shao:2018vul}.

In order to successfully implement the proposed $\dot\omega$/$\dot x_p$
tests, there are some concerns to address. Here we briefly recapitulate a
few key points~\cite{Shao:2018vul, Shao:2014bfa, Shao:2014oha}. (i) For pulsars whose $\dot x_p$ was not reported in 
literature, we conservatively estimate from the measured uncertainty of
$x_p$; the estimation was checked independently to be rather good with
PSRs~B1534+12 and B1913+16~\cite{Shao:2018vul}. (ii) The unknown $\Omega$ is treated as a {\it
nuisance} parameter in the Bayesian sense, and a randomization $\Omega \in
\left[ 0, 360^\circ \right)$ renders our tests as {\it probabilistic
tests}~\cite{Tanabashi:2018oca}. (iii) For binaries whose component masses
were derived from the accurately measured $\dot\omega$ {\it using GR}, we
recalculate them without using $\dot\omega$ (see 
Ref.\,\cite{Shao:2018vul} for discussions); therefore, we construct
``clean'' $\dot\omega$-tests albeit with a much worse precision. (iv) We take
care of the caution that a large $\dot\omega$ renders the secular changes
nonconstant~\cite{Wex:2007ct}. (v) We handle the fact that a large
proper motion for nearby binary pulsars introduces a nonzero $\dot
x_p$~\cite{Kopeikin:1996ads}. (vi) A fiducial radius $R_{\rm NS} = 12
\,$km is used, regardless of the complication from the equation of state.

\bgroup
\def\arraystretch{1.25}
\begin{table}
    \caption{Constraints on the GWEP violation from binary pulsars. \label{tab:limits}}
      \begin{tabular}{p{2cm}p{0.7cm}p{3.4cm}p{1.9cm}} 
	\hline\hline
	Pulsar & Test & Expression & 1$\sigma$ limit $[{\rm km}^4]$ \\
	\hline
	J0348+0432 & $\dot x_p$ & $\left| 0.59 K_{\hata\hatc} - 0.81
	K_{\hatb\hatc}  \right|$ &  $<8.2\times10^1$ \\
	J0737$-$3039A & $\dot x_p$ & $\left| 0.13 K_{\hata\hatc} - 0.99
	K_{\hatb\hatc}  \right|$ & $<4.8\times10^3$ \\
	& $\dot \omega$ & $\left| {\cal K} - 0.07 K_{\hata\hatc} - 0.01 K_{\hatb\hatc}  \right|$ &
	$<1.2\times10^5$ \\
	J0751+1807 & $\dot x_p$ & $\left|0.11 K_{\hata\hatc} -0.99
	K_{\hatb\hatc}  \right|$ & $<3.3\times10^2$ \\
	J1012+5307 & $\dot x_p$ & $\left| 0.39 K_{\hata\hatc} + 0.92
	K_{\hatb\hatc}  \right|$ & $<7.4\times10^2$ \\
	B1534+12 & $\dot x_p$ &  $\left|0.24 K_{\hata\hatc} + 0.97
	K_{\hatb\hatc}  \right| $ & $<4.8\times10^2$ \\
	& $\dot\omega$ & $\left| {\cal K} + 0.42 K_{\hata\hatc} - 0.11 K_{\hatb\hatc}  \right| $
	& $<1.7\times10^6$ \\
	J1738+0333 & $\dot x_p$ & $\left| 0.91 K_{\hata\hatc} + 0.41
	K_{\hatb\hatc}  \right|$ &  $<1.3\times10^2$ \\
	J1802$-$2124 & $\dot x_p$ & $\left| 0.94 K_{\hata\hatc} - 0.35
	K_{\hatb\hatc}  \right|$ & $<1.5\times10^4$ \\
	J1909$-$3744 & $\dot x_p$ & $\left| K_{\hata\hatc} \right|$ &
	$<5.2\times10^3$ \\
	B1913+16 & $\dot x_p$ & $\left| 0.16 K_{\hata\hatc} - 0.99
	K_{\hatb\hatc}  \right|$ & $<1.3\times10^2$ \\
	 & $\dot\omega$ & $\left| {\cal K} + 1.8 K_{\hata\hatc} + 0.30 K_{\hatb\hatc}  \right|$ &
	 $<1.0\times10^5$ \\
	 J2043+1711 & $\dot x_p$ & $\left| 0.55 K_{\hata\hatc} - 0.84
	 K_{\hatb\hatc}  \right| $ & $<6.7\times10^4$ \\
	B2127+11C & $\dot x_p$ & $\left| 0.96 K_{\hata\hatc} + 0.29
	K_{\hatb\hatc}  \right| $ & $<6.7\times10^3$ \\
	& $\dot\omega$ & $\left| {\cal K} + 0.50 K_{\hata\hatc} -1.6 K_{\hatb\hatc}  \right| $ &
	$<1.7\times10^6$ \\
	\hline
    \end{tabular}
\end{table}
\egroup

\section{Results}
\label{sec:res}

\begin{figure}
  \centering
    \includegraphics[width=7cm]{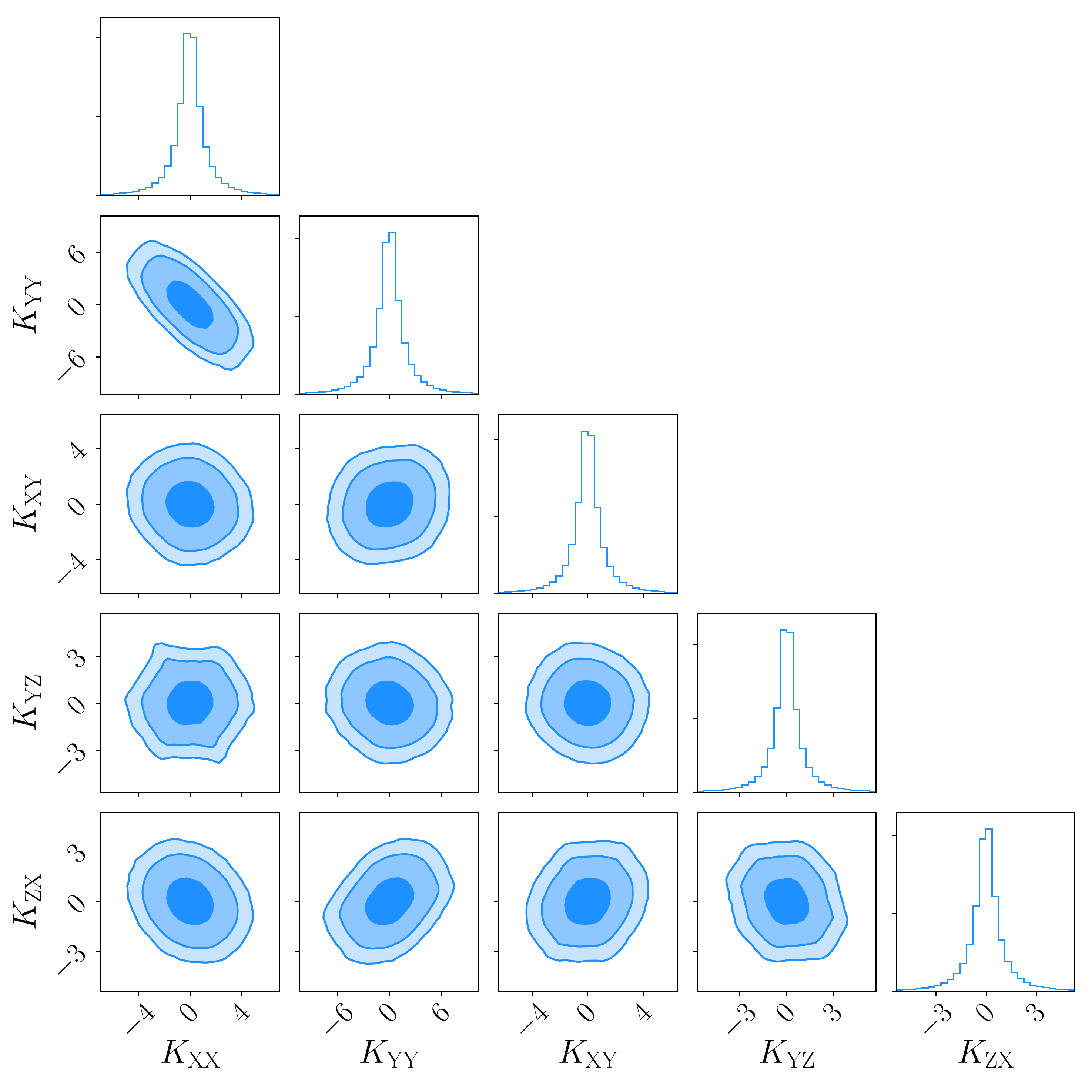}
    \caption{Constraints on $K_{jk}$ in the $\left( \hat{\bf X}, \hat{\bf Y},
    \hat{\bf Z} \right)$ frame. Contours show the 68\%, 90\%, and 95\% CLs.
    The unit for $K_{jk}$ is $10^3\,{\rm km}^4$. \label{fig:Kjk}}
  \end{figure}

After taking the above into account, we have derived 15
independent constraints in Table~\ref{tab:limits} on various linear
combinations of LV coefficients. The pulsars from the NS-WD
group provide one $\dot x_p$-test per system, and those from the NS-NS group
provide one $\dot x_p$-test and one $\dot\omega$-test per system. The limits
are in the range of ${\cal O}\left( 10^1 \, {\rm km}^4 \right)$ to ${\cal O}
\left( 10^6 \,{\rm km}^4 \right)$, in a broad agreement with the estimated
sensitivity~\cite{Bailey:2016ezm}. In general, the limits
from the $\dot\omega$-test are worse than those from the $\dot x_p$-test, due
to the large uncertainties in the {\it recalculated} masses for the sake of a
clean $\dot\omega$-test. The tightest limit, $\lesssim 10^2\,{\rm km}^4$,
comes from the $\dot x_p$ test of PSR~J0348+0432.

However, the limits in Table~\ref{tab:limits} are not expressed
in a common coordinate system. They
depend on the geometry of binary pulsars through the projections onto the
$\left( \hat{\bm{a}}, \hat{\bm{b}}, \hat{\bm{c}} \right)$ frame. Different
binaries carry different frames due to their different sky locations and
orbital orientations. To simplify comparisons with other experiments, 
it is standard to work in the Sun-centered celestial-equatorial coordinate system
$\left( \hat{\bf X}, \hat{\bf Y}, \hat{\bf Z}
\right)$~\cite{Kostelecky:2002hh, Kostelecky:2008ts}. As mentioned before,
$\left( \hat{\bm{a}}, \hat{\bm{b}}, \hat{\bm{c}} \right)$ and $\left(
\hat{\bf X}, \hat{\bf Y}, \hat{\bf Z} \right)$ frames are linked by a
rotation that composes of 5 simple parts related to the sky
location and orbital orientation; see Refs.~\cite{Shao:2014bfa, Shao:2018vul}
for details. We can relate the components of $K_{jk}$ in these two
frames~\cite{Bailey:2006fd} with e.g., $K_{\hata\hatb} = K_{jk} \hata^j
\hatb^k$ $(j,k = {\rm X,Y,Z})$. It allows us to express all limits in
Table~\ref{tab:limits} in terms of $K_{jk}$ in the $\left( \hat{\bf X},
\hat{\bf Y}, \hat{\bf Z} \right)$ frame and 5 rotation angles $\omega$,
$i$, $\Omega$, $\alpha$ (right ascension), and $\delta$ (declination). 

\def\arraystretch{1.25}
\begin{table}
    \caption{1$\sigma$ constraints on $K_{jk}$ in the
    $\left( \hat{\bf X}, \hat{\bf Y}, \hat{\bf Z} \right)$ frame. \label{tab:Kjk}}
      \begin{tabular}{p{2cm}p{3cm}p{3cm}}
	\hline\hline
	$K_{jk}$ & Scheme A $[{\rm km}^4]$ & Scheme B $[{\rm km}^4]$ \\
	\hline
	$\left|K_{\rm XX} \right|$ & $<1.6\times10^2$ & $<1.2\times10^3$ \\
	$\left|K_{\rm YY} \right|$ & $<2.4\times10^2$ & $<1.7\times10^3$ \\
	$\left|K_{\rm XY} \right|$ & $<1.8\times10^2$ & $<1.0\times10^3$ \\
	$\left|K_{\rm YZ} \right|$ & $<1.4\times10^2$ & $<8.6\times10^2$ \\
	$\left|K_{\rm ZX} \right|$ & $<1.6\times10^2$ & $<8.8\times10^2$ \\
	\hline
    \end{tabular}
\end{table}

\def\arraystretch{1.75}
\begin{table}
    \caption{1$\sigma$ constraints on the {\it absolute values} of $\bar{k}^{(8)}_{\rm eff}$ components  in Scheme C (unit: ${\rm km}^4$). Unconstrained ones are marked with dots. \label{tab:keff}}
      \begin{tabular}{p{1.7cm}p{2.6cm}p{1.7cm}p{1.8cm}}
  \hline\hline
  $\left|{\left(\bar{k}^{(8)}_{\rm eff}\right)}_{\rm XXXXXX} \right|$ & $< 2.3\times10^{2}$ & $\left|{\left(\bar{k}^{(8)}_{\rm eff}\right)}_{\rm XXXXXY} \right|$ & $< 1.8\times10^{2}$\\
$\left|{\left(\bar{k}^{(8)}_{\rm eff}\right)}_{\rm XXXXXZ} \right|$ & $< 1.6\times10^{2}$ & $\left|{\left(\bar{k}^{(8)}_{\rm eff}\right)}_{\rm XXXXYY} \right|$ & $< 5.1\times10^{2}$\\
$\left|{\left(\bar{k}^{(8)}_{\rm eff}\right)}_{\rm XXXXYZ} \right|$ & $< 1.4\times10^{2}$ & $\left|{\left(\bar{k}^{(8)}_{\rm eff}\right)}_{\rm XXXXZZ} \right|$ & $< 4.1\times10^{2}$\\
$\left|{\left(\bar{k}^{(8)}_{\rm eff}\right)}_{\rm XXXYXY} \right|$ & $< 1.7\times10^{2}$ & $\left|{\left(\bar{k}^{(8)}_{\rm eff}\right)}_{\rm XXXYXZ}\right|$ & $\cdots$\\
$\left|{\left(\bar{k}^{(8)}_{\rm eff}\right)}_{\rm XXXYYY} \right|$ & $< 2.7\times10^{2}$ & $\left|{\left(\bar{k}^{(8)}_{\rm eff}\right)}_{\rm XXXYYZ}\right|$ & $\cdots$\\
$\left|{\left(\bar{k}^{(8)}_{\rm eff}\right)}_{\rm XXXYZZ} \right|$ & $< 2.7\times10^{2}$ & $\left|{\left(\bar{k}^{(8)}_{\rm eff}\right)}_{\rm XXXZXZ} \right|$ & $< 1.7\times10^{2}$\\
$\left|{\left(\bar{k}^{(8)}_{\rm eff}\right)}_{\rm XXXZYY} \right|$ & $< 2.4\times10^{2}$ & $\left|{\left(\bar{k}^{(8)}_{\rm eff}\right)}_{\rm XXXZYZ}\right|$ & $\cdots$\\
$\left|{\left(\bar{k}^{(8)}_{\rm eff}\right)}_{\rm XXXZZZ} \right|$ & $< 2.4\times10^{2}$ & $\left|{\left(\bar{k}^{(8)}_{\rm eff}\right)}_{\rm XXYYYY} \right|$ & $< 3.6\times10^{2}$\\
$\left|{\left(\bar{k}^{(8)}_{\rm eff}\right)}_{\rm XXYYYZ} \right|$ & $< 2.1\times10^{2}$ & $\left|{\left(\bar{k}^{(8)}_{\rm eff}\right)}_{\rm XXYYZZ}\right|$ & $\cdots$\\
$\left|{\left(\bar{k}^{(8)}_{\rm eff}\right)}_{\rm XXYZYZ} \right|$ & $< 1.7\times10^{2}$ & $\left|{\left(\bar{k}^{(8)}_{\rm eff}\right)}_{\rm XXYZZZ} \right|$ & $< 2.1\times10^{2}$\\
$\left|{\left(\bar{k}^{(8)}_{\rm eff}\right)}_{\rm XXZZZZ} \right|$ & $< 2.8\times10^{2}$ & $\left|{\left(\bar{k}^{(8)}_{\rm eff}\right)}_{\rm XYXYXY} \right|$ & $< 1.4\times10^{2}$\\
$\left|{\left(\bar{k}^{(8)}_{\rm eff}\right)}_{\rm XYXYXZ} \right|$ & $< 1.2\times10^{2}$ & $\left|{\left(\bar{k}^{(8)}_{\rm eff}\right)}_{\rm XYXYYY} \right|$ & $< 2.2\times10^{2}$\\
$\left|{\left(\bar{k}^{(8)}_{\rm eff}\right)}_{\rm XYXYYZ} \right|$ & $< 1.1\times10^{2}$ & $\left|{\left(\bar{k}^{(8)}_{\rm eff}\right)}_{\rm XYXYZZ} \right|$ & $< 3.0\times10^{2}$\\
$\left|{\left(\bar{k}^{(8)}_{\rm eff}\right)}_{\rm XYXZXZ} \right|$ & $< 1.4\times10^{2}$ & $\left|{\left(\bar{k}^{(8)}_{\rm eff}\right)}_{\rm XYXZYY}\right|$ & $\cdots$\\
$\left|{\left(\bar{k}^{(8)}_{\rm eff}\right)}_{\rm XYXZYZ}\right|$ & $\cdots$ & $\left|{\left(\bar{k}^{(8)}_{\rm eff}\right)}_{\rm XYXZZZ}\right|$ & $\cdots$\\
$\left|{\left(\bar{k}^{(8)}_{\rm eff}\right)}_{\rm XYYYYY} \right|$ & $< 1.8\times10^{2}$ & $\left|{\left(\bar{k}^{(8)}_{\rm eff}\right)}_{\rm XYYYYZ}\right|$ & $\cdots$\\
$\left|{\left(\bar{k}^{(8)}_{\rm eff}\right)}_{\rm XYYYZZ} \right|$ & $< 2.7\times10^{2}$ & $\left|{\left(\bar{k}^{(8)}_{\rm eff}\right)}_{\rm XYYZYZ} \right|$ & $< 1.4\times10^{2}$\\
$\left|{\left(\bar{k}^{(8)}_{\rm eff}\right)}_{\rm XYYZZZ}\right|$ & $\cdots$ & $\left|{\left(\bar{k}^{(8)}_{\rm eff}\right)}_{\rm XYZZZZ} \right|$ & $< 1.8\times10^{2}$\\
$\left|{\left(\bar{k}^{(8)}_{\rm eff}\right)}_{\rm XZXZXZ} \right|$ & $< 1.2\times10^{2}$ & $\left|{\left(\bar{k}^{(8)}_{\rm eff}\right)}_{\rm XZXZYY} \right|$ & $< 2.2\times10^{2}$\\
$\left|{\left(\bar{k}^{(8)}_{\rm eff}\right)}_{\rm XZXZYZ} \right|$ & $< 1.1\times10^{2}$ & $\left|{\left(\bar{k}^{(8)}_{\rm eff}\right)}_{\rm XZXZZZ} \right|$ & $< 3.0\times10^{2}$\\
$\left|{\left(\bar{k}^{(8)}_{\rm eff}\right)}_{\rm XZYYYY} \right|$ & $< 1.6\times10^{2}$ & $\left|{\left(\bar{k}^{(8)}_{\rm eff}\right)}_{\rm XZYYYZ}\right|$ & $\cdots$\\
$\left|{\left(\bar{k}^{(8)}_{\rm eff}\right)}_{\rm XZYYZZ} \right|$ & $< 2.4\times10^{2}$ & $\left|{\left(\bar{k}^{(8)}_{\rm eff}\right)}_{\rm XZYZYZ} \right|$ & $< 1.2\times10^{2}$\\
$\left|{\left(\bar{k}^{(8)}_{\rm eff}\right)}_{\rm XZYZZZ}\right|$ & $\cdots$ & $\left|{\left(\bar{k}^{(8)}_{\rm eff}\right)}_{\rm XZZZZZ} \right|$ & $< 1.6\times10^{2}$\\
$\left|{\left(\bar{k}^{(8)}_{\rm eff}\right)}_{\rm YYYYYY} \right|$ & $< 2.9\times10^{2}$ & $\left|{\left(\bar{k}^{(8)}_{\rm eff}\right)}_{\rm YYYYYZ} \right|$ & $< 1.4\times10^{2}$\\
$\left|{\left(\bar{k}^{(8)}_{\rm eff}\right)}_{\rm YYYYZZ} \right|$ & $< 2.9\times10^{2}$ & $\left|{\left(\bar{k}^{(8)}_{\rm eff}\right)}_{\rm YYYZYZ} \right|$ & $< 2.2\times10^{2}$\\
$\left|{\left(\bar{k}^{(8)}_{\rm eff}\right)}_{\rm YYYZZZ} \right|$ & $< 2.1\times10^{2}$ & $\left|{\left(\bar{k}^{(8)}_{\rm eff}\right)}_{\rm YYZZZZ} \right|$ & $< 2.5\times10^{2}$\\
$\left|{\left(\bar{k}^{(8)}_{\rm eff}\right)}_{\rm YZYZYZ} \right|$ & $< 1.1\times10^{2}$ & $\left|{\left(\bar{k}^{(8)}_{\rm eff}\right)}_{\rm YZYZZZ} \right|$ & $< 3.0\times10^{2}$\\
$\left|{\left(\bar{k}^{(8)}_{\rm eff}\right)}_{\rm YZZZZZ} \right|$ & $< 1.4\times10^{2}$ & $\left|{\left(\bar{k}^{(8)}_{\rm eff}\right)}_{\rm ZZZZZZ} \right|$ & $< 4.0\times10^{2}$\\
	\hline
    \end{tabular}
\end{table}

To proceed practically, for a 1$\sigma$ limit ``$a$'' in
Table~\ref{tab:limits}, denoted as $\left| {\cal X}_a \left( K_{jk}, \Omega_a \right)
\right| < {\cal C}_a$ (e.g., $\left| 0.59 K_{\hat a \hat c} - 0.81
K_{\hat b \hat c} \right| < 8.2 \times 10^1\,{\rm km}^4$ from
PSR~J0348+0432), we use the following probabilistic density function
(PDF) \cite{Shao:2018vul},
\begin{align}
  P\left( K_{jk} \right) \propto \prod_a \int_0^{2\pi} \exp \left\{
  -\frac{1}{2} \left| \frac{ {\cal X}_a \left( K_{jk},\Omega_a \right) }{
  {\cal C}_a} \right|^2 \right\} \rmd \Omega_a \,,
  \label{eq:pdf}
\end{align}
where we have made assumptions on the Gaussianity of measurements and the
independence of the limits in Table~\ref{tab:limits}. 

We use three schemes to obtain sensible limits on $K_{jk}$. In scheme~A, we
make an assumption that only one component of $K_{jk}$ in the $\left(
\hat{\bf X}, \hat{\bf Y}, \hat{\bf Z} \right)$ frame is nonzero. We obtain
PDF for each component, and extract the limits at 68\% CL (see
Table~\ref{tab:Kjk}). The limits are dominated by the tightest limit in
Table~\ref{tab:limits}. In scheme~B, we allow all components of $K_{jk}$ to
be nonzero. We use the {\sc emcee} package~\cite{ForemanMackey:2012ig} to
investigate their joint PDF with Markov-chain Monte Carlo (MCMC) simulations.
We use 20 walkers to accumulate $10^7$ samples in total, of which the first
half is discarded as the {\sc burn-in} phase. The 2d pairwise distributions
for $K_{jk}$ are shown in Figure~\ref{fig:Kjk}, together with 1d marginalized
distributions. We extract the limits for each component at 68\% CL, tabulated
in Table~\ref{tab:Kjk}. We can see that the limits are only worse than those
from scheme~A by a factor of a few. We have 5 degrees of freedom and 15
constraints. The overconstraining system bounds the values of these 5
components very efficiently~\cite{Shao:2014oha}. In scheme~C, we assume only
one component of $\bar k^{(8)}_{\rm eff}$ in the $\left( \hat{\bf X},
\hat{\bf Y}, \hat{\bf Z} \right)$ frame is nonzero; results are given in
Table~\ref{tab:keff}. In this case, we can constrain 45 components of $\bar
k^{(8)}_{\rm eff}$; the other 11 components do not appear in $K_{jk}$.

The limits in Tables~\ref{tab:Kjk} and~\ref{tab:keff} are the first sets of
this kind on the nonlinear terms and body-dependent effects in the gravity
sector of the SME. We suggest that other groups can perform similar analysis
on their experiments and compare the results with ours. In short-range
experiments, a different treatment is needed, and they might probe unconstrained components in this study. We do not
make efforts to translate from $K_{jk}$ and $\bar k^{(8)}_{\rm eff}$ to $
\bar k^{(8)}_{\cal ABC}$, because $K_{jk}$ and $\bar k^{(8)}_{\rm eff}$ have
the actual parameters which appear in the relative acceleration
(\ref{eq:del:relative:acc}) and the Poisson-like equation (\ref{eq:deltah00})
respectively. It is universal for a variety of experiments, and easy for
future comparisons.

\section{Discussions}
\label{sec:diss}

In this {\it Letter} we follow the theoretical work by~\citet{Bailey:2016ezm}
and investigate in detail the GWEP-violating signals in the gravity sector of
the SME with binary pulsars. Nonlinear terms from the anisotropic cubic
curvature couplings (\ref{eq:L8}) introduce novel effects that are absent in
the linearized gravity. Most notably, the relative accleration between two
objects depends on the internal structure of the bodies. The denser the
object (or, the more compact the object), the larger the abnormal
acceleration. NSs are among the densest objects, hence they are intrinsically
privileged in testing this kind of GWEP-violating phenomenon. We use multiple
pulsars that were prepared to test the CPT-violating gravity in the
SME~\cite{Shao:2018vul}, and put the first sets of bounds on relevant
parameters. Our bounds on the GWEP-violating parameters are listed in
Tables~\ref{tab:Kjk} and \ref{tab:keff}. We hope other experiments will
provide complementary bounds, probably on different degrees of freedom.

The violation of GWEP reminds us the famous ``Nordtvedt
effect''~\cite{Nordtvedt:1968qs}. It is one of the main ingredients for the
strong equivalence principle (SEP)~\cite{Will:1993ns, Will:2014kxa}. For all
the valid alternative gravity theories, SEP basically implies the
uniqueness of GR~\cite{Will:2014kxa}. Therefore, the tests of the GWEP in
this work are important. It provides complementary information to the
existing tests of the GWEP in specific alternative gravity theories, like 
the scalar-tensor gravity~\cite{Barausse:2017gip}.

It is worthy to mention that, in the theoretical
treatment~\cite{Bailey:2016ezm}, we have only used the quadratic
${\cal O}\left( h^2 \right)$ modifications in the field equation,
neglecting all the other higher-order terms.  In this sense, our
limits should be treated as the strong-field {\it effective} limits to
their weak-field counterparts; see Refs.~\cite{Will:2018ont,
Shao:2012eg, Shao:2013wga} for more details. Already at this
approximation, we begin to obtain body-dependent GWEP-violating
effects.  By fully incorporating all the nonlinear terms might provide
even more interesting phenomenona. In a class of scalar-tensor
gravity, nonperturbative strong-field effects were discovered with
fully nonlinear equations~\cite{Damour:1993hw, Damour:1996ke,
Shao:2017gwu, Sennett:2017lcx}, that provided important tests for
gravitation in the strong-field regime~\cite{Will:2014kxa,
Wex:2014nva}. Possible extension of the gravity sector in the SME to
higher orders is beyond the scope of this work.

From the perspective of pulsar timing, continuous observations will
improve the $\dot\omega$ and $\dot x_p$ accuracy as $T^{-3/2}$, where
$T$ is the observational time span.  Thus the $T^{-3/2}$ improvement
for $\dot\omega$/$\dot x_p$ tests is guaranteed. The upcoming large
radio telescopes and arrays will further tighten the bounds. For
example, the Five-hundred-meter Aperture Spherical Telescope (FAST)
\cite{Nan:2011um} and the MeerKAT array \cite{Bailes:2018azh} are
starting to operate, and will provide a big improvement in the timing
precision. Ultimately for the next decades, the Square Kilometre Array
\cite{Kramer:2004hd, Shao:2014wja, Bull:2018lat}, is going to test
gravity in an unparalleled way with its remarkable sensitivity.

\acknowledgments

We are grateful to Alan Kosteleck\'y and Norbert Wex for helpful discussions, 
and three anonymous referees for their comments and critiques.
This work was supported by the Young Elite Scientists Sponsorship Program by
the China Association for Science and Technology (2018QNRC001),
and partially supported by the National Science Foundation of China (11721303),
XDB23010200, and the European Research Council (ERC) for the ERC Synergy Grant
BlackHoleCam under Contract No. 610058. QGB acknowledges the National Science
Foundation of the USA for support under grant number PHY-1806871.


%

\end{document}